\documentstyle[aps,multicol,prb,epsf]{revtex}
\begin{document}
\title{Kondo effect in complex mesoscopic structures}
\author{O. Entin-Wohlman$^{a,b,c}$, A. Aharony$^{a,b}$, and Y. Meir$^{a,d}$}
\address{$^a$Department of Physics, Ben Gurion University, Beer
Sheva 84105, Israel} \address{$^b$ School of Physics and
Astronomy, Raymond and Beverly Sackler Faculty of Exact
Sciences,\\
Tel Aviv University, Tel Aviv 69978, Israel}
\address{$^c$Albert
Einstein Minerva Center for Theoretical Physics, Weizmann
Institute of Science, Rehovot 76100, Israel}
\address{The Ilse Katz Center for Meso- and Nano-Scale
Science and Technology, Ben Gurion University, Beer Sheva 84105,
Israel}
\date{\today}
\maketitle

\begin{abstract}
We study the Kondo effect of a quantum dot placed in a complex
mesoscopic structure. Assuming that electronic interactions are
taking place solely on the dot, and focusing on the infinite
Hubbard interaction limit, we use a decoupling scheme to obtain an
explicit analytic approximate expression for the dot Green
function, which fulfills certain Fermi-liquid relations at zero
temperature. The details of the complex structure enter into this
expression only via the self-energy for the non-interacting case.
The effectiveness of the expression is demonstrated for the single
impurity Anderson model and for the T-shaped network.
\end{abstract}

\begin{multicols}{2}

\section{introduction and outline}

The single-impurity Anderson model (SIAM) has been a paradigm of a
strongly correlated electron system. \cite{fulde} This seemingly
simple model gives rise to dynamical screening of the local spin
by the electrons in the Fermi sea, leading to a crossover from a
weak coupling system at high temperatures ($T$) to a strongly
coupled one at low $T$, with the relevant temperature scale given
by the Kondo temperature, $T_K$. The recent observation of the
Kondo effect in quantum dots (QD's), whose parameters can be tuned
continuously, and which allow probing of various properties,
\cite{mit} has yielded strong theoretical efforts in this
direction. \cite{wingreen} Recent experiments on a QD embedded on
one branch of the Aharonov-Bohm interferometer \cite{ji} created
additional interest in Kondo effects in {\it complex networks}.

While the high-$T$ behavior of the SIAM can be adequately
described by perturbation theory or poor-man scaling, \cite{sivan}
and the low-$T$ behavior is described by Fermi liquid theory,
\cite{nozieres} there is no {\it simple} theory that describes the
model's dynamical  properties correctly in the whole temperature
range, including {\it both} the high- and the low-$T$ limits. This
crossover has been described by the computationally demanding
numerical renormalization group (NRG) approach. \cite{costi} Other
methods to describe the crossover are the ``conserving T-matrix
approximation" \cite{wolfle} (which overestimates the unitarity
sum-rule), or the more limited ``non-crossing approximation",
\cite{bickers} and to some extent, quantum Monte Carlo
calculations. \cite{hirsch} Some of these methods have also been
applied to the QD in the Aharonov-Bohm interferometer.
\cite{ABNRG} However, none of these methods has the flexibility to
follow {\it analytically} the effects of various network
parameters (e.g. a magnetic flux) on experimentally measurable
quantities, in particular dynamical ones.

In this paper we discuss the SIAM for a QD which is embedded in a
{\it general complex network}. Many of the interesting physical
properties of the system can be expressed explicitly in terms of
the single electron (retarded) Green function (GF) on the dot,
$G_{dd\sigma}(\omega)$, for electrons with spin component $\sigma$
and energy $\omega$ (measured relative to the Fermi energy). Here,
\begin{eqnarray}
G_{AB\sigma}(\omega)&&\equiv\ll A_{\sigma};B^{\dagger}_{\sigma}\gg
\nonumber\\
&&\equiv -i\int_{0}^{\infty}dt e^{i(\omega +i\eta )t}\langle
[A_{\sigma}(t),B^\dagger_{\sigma}]_+\rangle.
\end{eqnarray}
(For the dot GF, $A=B=d_{\sigma}$, where $d_{\sigma}$ destroys an
electron with spin $\sigma$ on the dot).  One example concerns the
local density of states (LDOS) on the dot,
\begin{eqnarray}
\rho_{d\sigma}(\omega)=-\Im [G_{dd\sigma}(\omega)]/\pi.\label{rho}
\end{eqnarray}
Another example is the conductance ${\cal G}$ between two leads
which are connected to sites on the network. For a general
mesoscopic structure,  this (linear response) conductance has the
form \cite{we}
\begin{eqnarray}
{\cal G}=\frac{2e^2}{h}\int d\omega \Bigl (-\frac{\partial f
}{\partial \omega}\Bigr ){\tilde {\cal G}}(\omega),\label{cond}
\end{eqnarray}
where $f(\omega )$ is the Fermi distribution function (we set the
Fermi energy at $\omega=0$),
\begin{eqnarray}
f(\omega )=\frac{1}{e^{\omega/kT }+1},\label{FER}
\end{eqnarray}
and ${\tilde {\cal G}}(\omega)$ is often given in terms of the
retarded $G_{dd\sigma}$, the advanced $G^\ast_{dd\sigma}$ and
non-interacting parameters characterizing the network. For the
simple SIAM (dot and leads), ${\tilde {\cal G}}(\omega)$ is
proportional to $\rho_{d\sigma}(\omega)$. \cite{ng,meir}

Here we use the equations of motion  method to derive a simple
analytic approximate expression for $G_{dd\sigma}$, for a QD on a
general network. Our approximate formula exhibits the correct
behavior both at high and at very low temperatures. For
simplicity, we consider a QD with a single level (of energy
$\epsilon_{d}$). Electron-electron interactions are assumed to
exist only on the QD, and we take their energy $U$ to be infinite.
In our scheme, one first solves for the GF on the dot {\it in the
absence of the electron-electron interactions}, i.e. $U=0$. This
(spin-independent) GF can be written in the form
\begin{eqnarray}
G^0_{dd}(\omega)=\frac{1}{\omega+i\eta-\epsilon_d-\Sigma_0(\omega)},
\label{G0}
\end{eqnarray}
with the energy dependent (complex)
self-energy (SE)
\begin{eqnarray}
\Sigma_{0}(\omega)\equiv \delta\epsilon_d(\omega)-i\Delta(\omega).
\label{ris}
\end{eqnarray}
This is easily done exactly, for any finite network; it only
involves the solution of a finite set of linear equations. Our
approximate explicit expression for $G_{dd\sigma}(\omega)$ is then
given in a single equation [see Eq. (\ref{GGG}) below]. Somewhat
surprisingly, this equation depends on the network parameters {\it
only} via $\Sigma_{0}(\omega)$. Although approximate, this
expression allows for detailed systematic investigations of the
Kondo effect as function of the system parameters, on a variety of
complex networks. Even if sometimes only qualitatively correct,
such systematic studies help to investigate new physical phenomena
on a broad variety of mesoscopic systems. Furthermore, the
dynamical mean field theory \cite{mf} (which was developed to
address the physics of the periodic Anderson and Hubbard models)
iterates the local GF of the SIAM, which is calculated at each
stage in terms of the effective SE created by the rest of the
lattice. Since our GF is easily calculated in terms of $\Sigma_0$,
it is an ideal candidate for such calculations. We are not aware
of alternative simple analytic expressions which obey the
necessary requirements at both high and low $T$.

In Sec. \ref{II} we follow Refs.
\onlinecite{dworin,penn,lacroix,luo}, and derive the
equation-of-motion (EOM) for the dot GF. Extending these
references, this EOM is derived here for a {\it generalized case},
in which the dot ``sits" within an {\it arbitrary complex
network}.
 The EOM for the dot GF
involves higher-order GF's (consisting of more operators), whose
EOM's introduce in turn more GF's. One then terminates this
hierarchy by decoupling out averages of operators. The latter are
then found using the fluctuation-dissipation relationship, from
the relevant GF's. In this manner, the treatment becomes
self-consistent. The successful decoupling must keep as much of
the electronic correlations as possible. \cite{penn} For example,
a widely used earlier approximation \cite{meir} neglected some
dot-lead correlations and therefore gave reasonable results {\it
only} at $T>T_K$. For this reason, Gerland {\it et al.}
\cite{oreg} had to combine the EOM method at high $T$ with the NRG
at low $T$. After we correct these earlier calculations, and
include all the necessary correlations, we obtain good results for
{\it all} $T$. An earlier application of the EOM method solved the
integral equations numerically, on the simplest interferometer
geometry.
 \cite{bulka}

The above decoupling scheme produces an integral equation for
$G_{dd\sigma}(\omega)$, which generalizes that found in Refs.
\onlinecite{lacroix} and \onlinecite{bloomfield}.  A new
approximate analytical solution for this equation is found in Sec.
\ref{III}. The low temperature limit, the Fermi liquid conditions,
and the Kondo behavior are discussed in Sec. \ref{fermi}. Unlike
earlier papers (spanning over four decades), our solution gives
good qualitative results for the {\it whole parameter range}: It
has the correct high-$T$ behavior and fulfills several low-$T$
Fermi liquid relations.

For the simplest SIAM, the ``network" is represented by a band
with a density of states $N(\omega)$. In the broad band limit, the
self-energy on the QD can be approximated by its value at the
Fermi energy, $\Sigma_0(\omega) \approx -i \Delta_b=-i \pi V^2
N(0)$, and $\Delta_b$ represents the width of the impurity state
in the absence of interactions ($V$ represents the approximately
energy-independent dot-band coupling). In this case, our equations
reduce to those discussed in the earlier literature. We emphasize
again that even the solution for this simple case is new and
non-trivial, since earlier work either did not have a close
analytical formula or missed some correlations, leading to wrong
results at low $T$.

As stated, our result for $G_{dd\sigma}(\omega)$ depends on the
details of the general network only via the non-interacting
self-energy, $\Sigma_0(\omega)$. To demonstrate the power of our
result, we present here a few simple examples. Section \ref{IV}
contains explicit results for the simple case of the SIAM, when
the QD is coupled to two one-dimensional leads.  For this case, we
easily calculate various features of the Kondo behavior, including
the peak in $\rho_{dd\sigma}$ at the Fermi energy, the plateau in
the conductance in the so-called unitary limit and the plateau in
the ``transmission phase" at $\pi/2$.

A second example, of the so called ``T" network, is solved in Sec.
\ref{T}. In this case, one has interference between the wave
function on the dot and that on the intersection point. Our simple
approximate formula exhibits the Fano vanishing of the
transmission and the associated so-called anti-Kondo effect, as
seen experimentally. \cite{Texp} Here we also extend earlier
theoretical work, which used complicated techniques.
\cite{Ttheor,jps} Finally, Sec. \ref{V} contains our summary.

\section{Equations of Motion}\label{II}

\subsection{The Hamiltonian}

Our Hamiltonian,
\begin{eqnarray}
{\cal H}={\cal H}_{\rm dot}+{\cal H}_{\rm net}+{\cal H}_{\rm
dot-net}, \label{H1}\end{eqnarray}
 contains the dot part
 \begin{eqnarray}
 {\cal H}_{d}=\epsilon_{d}\sum_{\sigma}n_{d\sigma}+
 Un_{d\uparrow}n_{d\downarrow},\label{H2}
 \end{eqnarray}
with a single energy level $\epsilon_d$ and an electron-electron
interaction $U$ ($n_{d\sigma}=d^{\dagger}_{\sigma}d_{\sigma}$).
All the other parts of ${\cal H}$ can be arbitrary, but
non-interacting:
\begin{eqnarray}
{\cal H}_{\rm net}=\sum_{n\sigma}
\epsilon_{n}a_{n\sigma}^{\dagger}a_{n\sigma}
-\sum_{nm\sigma}J_{nm}a^{\dagger}_{n\sigma}a_{m\sigma}\label{H3}
\end{eqnarray}
 describes the network without the dot, and
 \begin{eqnarray}
 {\cal H}_{\rm
dot-net}=-\sum_{n\sigma}\bigl (J_{dn}d_\sigma^\dagger
a_{n\sigma}+h.c. \bigr )\label{H4}
\end{eqnarray}
describes the coupling of the dot to the network. Here,
$a_{n\sigma}^{\dagger}$ creates an electron with spin $\sigma$ and
energy $\epsilon_n$ on the site $n$ of the network. The
coefficients $J_{nm}$ and $J_{dn}$ represent tight-binding hopping
matrix elements. In the absence of a magnetic flux, these
coefficients can be chosen to be real. The flux then turns them
complex, with phases which relate to the Aharonov-Bohm effect and
with $J_{nm}=J_{mn}^\ast$. \cite{ABNRG,bulka,ABphases}

The main part of this paper concerns the above general
Hamiltonian, which assumes no specific details on the structure of
the network. However, for some specific applications, it is
convenient to attach a finite mesoscopic network to several
semi-infinite leads, which connect to electron reservoirs. We thus
divide ${\cal H}_{\rm net}$ into two parts,
\begin{eqnarray}
{\cal H}_{\rm net}={\cal H}_{\rm meso}+{\cal H}_{\rm leads},
\end{eqnarray}
where ${\cal H}_{\rm meso}$ has the same form as ${\cal H}_{\rm
net}$, except that the sum is restricted to sites which belong to
the finite mesoscopic part of the net (excluding the dot and the
leads), and
\begin{eqnarray}
{\cal H}_{\rm leads}&=&\sum_L\bigl [{\cal H}_L-\sum_{n_L,r}\bigl
(J_{n_L,r}a^\dagger_{n_L\sigma}c^{(L)}_{r\sigma}
+h.c.\bigr )\bigr ],\nonumber\\
 {\cal H}_L&=&-\sum_{rs} J_{rs}
c^{(L)\dagger}_{r\sigma} c^{(L)}_{s\sigma}+h.c..\label{leads}
\end{eqnarray}
Here $n_L$ are the indices of the points on the mesoscopic
structure which are connected to lead number $L$, while
$c^{(L)}_{r\sigma}$ destroys an electron of spin $\sigma$ on the
site $r$ of the lead.

Usually, the lead $L$
 has a continuous spectrum, with eigenstates $|k\rangle$
and eigenenergies $\epsilon_k$, in a band of width $2D$. Equation
(\ref{leads}) can then be written as
\begin{eqnarray}
{\cal H}_{\rm leads}&=&\sum_L\bigl [{\cal H}_L-\sum_{n_L,k}\bigl
(V_{n_L}(k)a^\dagger_{n_L\sigma}c^{(L)}_{k\sigma}
+h.c.\bigr )\bigr ],\nonumber\\
 {\cal H}_L&=&\sum_{k}\epsilon_k
c^{(L)\dagger}_{k\sigma} c^{(L)}_{k\sigma},\label{leads1}
\end{eqnarray}
and also
\begin{eqnarray}
 {\cal H}_{\rm
dot-net}&=&-\sum_{n\sigma}\bigl (J_{dn}d_\sigma^\dagger
a_{n\sigma}\nonumber\\
&-&\sum_L\sum_{k\sigma}V_d(k)d^\dagger_\sigma c^{(L)}_{k\sigma}
+h.c. \bigr ),
\end{eqnarray}
with obvious definitions. When the  network consists only of the
dot and the leads (i.e. ${\cal H}_{\rm meso}=0$) then this has
exactly the form of the standard SIAM model, discussed broadly in
the literature. If all the leads have similar bands, then the
electrons in the whole system also have a continuous spectrum in
the range $-D<\omega<D$. All integrals over $\omega$ below will be
thus over this energy band, and it is due to resonances in such
integrals that we need to calculate the retarded GF at
$\omega+i\eta$.

\subsection{Derivation of $\Sigma_0(\omega)$}

From now on we return to the general Hamiltonian, Eqs.
(\ref{H1}-\ref{H4}). The EOM for the retarded GF on the QD reads
\begin{eqnarray}
(\omega+i \eta -\epsilon_{d})G_{dd\sigma}&=&1 -\sum_n
J_{dn}G_{nd\sigma} +U\Gamma_{\sigma}, \label{eomG0}
\end{eqnarray}
where \begin{eqnarray} \Gamma_{\sigma}=\ll
n_{d-\sigma}d_{\sigma};d^{\dagger}_{\sigma}\gg. \end{eqnarray}
From now on, we shall use the shorthand $\omega$ for
$\omega+i\eta$, except where it matters. Specifically,
\begin{eqnarray}
(\omega -\epsilon_n)G_{nd\sigma}=-J^\ast_{dn}G_{dd\sigma}-\sum_m
J_{nm}G_{md\sigma}. \label{Gnd}
\end{eqnarray}
We next define the matrix
\begin{eqnarray}
{\cal M}(\omega)_{nm} \equiv \langle n|\omega-{\cal
H}_{\rm net}|m\rangle
\equiv J_{nm}+\delta_{nm}(\omega-\epsilon_n),
\end{eqnarray}
which represents the system without the dot. In the following, we
shall need the inverse matrix, \begin{eqnarray} {\cal F}(\omega)
\equiv [{\cal M}(\omega)]^{-1}, \end{eqnarray} which is the Green
function for the system without the dot.
With this definition one
finds
\begin{eqnarray}
G_{md\sigma}=u_m(\omega) G_{dd\sigma}, \label{Gmd}
\end{eqnarray}
where
\begin{eqnarray}
u_m(\omega) \equiv -\sum_n{\cal F}(\omega)_{mn}J_{nd}.\label{um}
\end{eqnarray}
We can now calculate the second term on the RHS of Eq.
(\ref{eomG0}), and show that
\begin{eqnarray}
-\sum_n J_{dn}G_{nd\sigma}=\Sigma_{0}G_{dd\sigma},\label{sigext}
\end{eqnarray}
with
\begin{eqnarray}
\Sigma_0(\omega) \equiv -\sum_m J_{dm}u_m(\omega)
\equiv \sum_{mn}J_{dm}{\cal F}(\omega)_{mn}J_{nd}. \label{sig0}
\end{eqnarray}
At $U=0$, Eq. (\ref{eomG0}) thus reduces to Eq. (\ref{G0}).
Clearly, $\Sigma_0$
 can be calculated for the
{\it non-interacting} case.

All the other GF's can similarly be expressed in terms of
$G_{dd\sigma}$. For example, \begin{eqnarray}
G_{dm\sigma}=u_m^\ast(\omega^\ast)G_{dd\sigma},\label{Gdm}
\end{eqnarray}
where $\omega^\ast \equiv \omega-i\eta$. Another way to represent
$u_m^\ast(\omega^\ast)$ is to start from $u_m(\omega)$, and then
take the complex conjugate only of the hopping coefficients
$J_{mn}$ (namely reverse all their phases), without touching
$\omega+i\eta$. Similarly,
\begin{eqnarray}
G_{\ell m\sigma}={\cal F}(\omega)_{\ell m}+u_\ell(\omega)
G_{dm\sigma}.\label{Glm}
\end{eqnarray}

Since ${\cal M}$ is an infinite matrix, its inversion may not be
trivial. Formally, we can denote the eigenenergies and eigenstates
of ${\cal H}_{\rm net}$ by $\epsilon_a$ and $|a \rangle$, and then
write
\begin{eqnarray}
{\cal F}(\omega)_{mn}=\sum_a \langle m|a\rangle
\frac{1}{\omega+i\eta-\epsilon_a}\langle a|n\rangle. \label{MM}
\end{eqnarray}
However, in the examples it is useful first to eliminate the parts
which involve the sites on the leads. Examples for this procedure
are presented in Appendix \ref{A}.

\subsection{EOM's for higher-order GF's}

The EOM for $\Gamma_{\sigma}$ (with only on-site Hubbard
interaction on the dot) does not involve GF's of more arguments:
\begin{eqnarray}
(\omega -\epsilon_{d}-U)\Gamma_{\sigma}&=&\langle
n_{d-\sigma}\rangle +\sum_n J_{dn}^\ast \Gamma^{(3)}_{n\sigma}
\nonumber\\
&-&\sum_{n} J_{dn}(\Gamma_{n\sigma}^{(1)}
+\Gamma_{n\sigma}^{(2)}), \label{eomgamma}
\end{eqnarray}
with three new GF's, \begin{eqnarray} \Gamma^{(1)}_{n\sigma}&=&\ll
n_{d-\sigma}a_{n\sigma};d^{\dagger}_{\sigma}\gg ,
\nonumber\\
\Gamma^{(2)}_{n\sigma}&=&\ll d^{\dagger}_{-\sigma}a_{n
-\sigma}d_{\sigma};d^{\dagger}_{\sigma}\gg ,
\nonumber\\
\Gamma^{(3)}_{n\sigma}&=&\ll a^{\dagger}_{n
-\sigma}d_{-\sigma}d_{\sigma};d^{\dagger}_{\sigma}\gg.
\end{eqnarray}
The inhomogeneous term in Eq. (\ref{eomgamma}), $\langle
n_{d-\sigma} \rangle$, represents the average number of electrons
with spin $\sigma$ on the dot. This number needs to be determined
self-consistently, via the fluctuation-dissipation theorem:
\begin{eqnarray}
\langle n_{d\sigma} \rangle =\int d\omega f(\omega)
\rho_{d\sigma}(\omega). \label{selfcon} \end{eqnarray}
 We shall
return to this condition below.

We now write the EOM's for the new GF's, $\Gamma^{(i)}_{n\sigma}$,
and use several approximations for their solution. Firstly, it is
easy to see from the EOM for $\Gamma^{(3)}_{n\sigma}$ that this GF
is of order $1/U$. Therefore, it contributes to $\Gamma_\sigma$
only at order $1/U^2$, and can be ignored for $U \rightarrow
\infty$ [we need $\Gamma_\sigma$ only to order $1/U$, see Eq.
(\ref{eomG0})]. Secondly, we introduce a decoupling scheme for the
new GF's that appear in the EOM's of the
$\Gamma^{(i)}_{n\sigma}$'s,
\begin{eqnarray}
\ll
a^{\dagger}_{n\sigma_{1}}
a_{m\sigma_{2}}d_{\sigma_{3}}&;&d^{\dagger}_{\sigma}\gg
\simeq \langle
a^{\dagger}_{n\sigma_{1}}
a_{m\sigma_{1}}\rangle\delta_{\sigma_{1}\sigma_{2}}\ll
d_{\sigma_{3}};d^{\dagger}_{\sigma}\gg \nonumber\\
 &-&\langle
a^{\dagger}_{n\sigma_{1}}
d_{\sigma_{1}}\rangle\delta_{\sigma_{1}\sigma_{3}}\ll
a_{m\sigma_{2}};d^{\dagger}_{\sigma}\gg .
\end{eqnarray}
We omit thermal averages of the form $\langle
a_{n\sigma}d_{\sigma}\rangle$, which include two destruction (or
creation) operators, and $\langle
a^{\dagger}_{n\sigma}a_{m\sigma'}\rangle$ with $\sigma \ne
\sigma'$, relevant only for states with a net magnetic moment. The
latter assumption means that we restrict the discussion only to
symmetric states, with $G_{dd\uparrow}=G_{dd\downarrow}$.

Using these approximations,  the EOM for $\Gamma_{n\sigma}^{(1)}$
becomes
\begin{eqnarray}
(\omega-\epsilon_n)\Gamma_{n\sigma}^{(1)}&=&- J_{dn}^\ast
\Gamma_\sigma-\sum_m J_{nm}\Gamma_{m\sigma}^{(1)}. \label{G1}
\end{eqnarray}
The derivation of this equation also required an additional term,
$-G_{nd\sigma}[\Sigma_m J_{dm} \langle d^\dagger_{-\sigma}
a_{m-\sigma} \rangle-c.c.]$. However, this term vanishes. Here and
below, we calculate equilibrium thermal averages by using the
fluctuation-dissipation theorem,
\begin{eqnarray}
\langle A^{\dagger}B\rangle =\int \Bigl (-\frac{d\omega}{2\pi
i}\Bigr )f(\omega )\Bigl (G_{BA}-G^{\ast}_{AB}\Bigr ).\label{FD}
\end{eqnarray}

 Equation (\ref{G1}) is practically the same as
Eq. (\ref{Gnd}), which was used to derive $\Sigma_0$ in terms of
the $J_{nm}$'s. Applying the same algebra to Eq. (\ref{G1}) then
yields the analog to Eq. (\ref{sigext}),
\begin{eqnarray}
-\sum_n J_{dn}
\Gamma_{n\sigma}^{(1)}=\Sigma_{0}\Gamma_{\sigma}.\label{g1sof}
\end{eqnarray}

Using the same approximations, the EOM of $\Gamma_{n\sigma}^{(2)}$
is
\begin{eqnarray}
(\omega &-&\epsilon_n)\Gamma_{n\sigma}^{(2)}=-J_{dn}^\ast
\Gamma_{\sigma}-\sum_m J_{nm}\Gamma_{m\sigma}^{(2)}\nonumber\\
&+&\langle X_{n}\rangle G_{dd\sigma}+\langle
d^{\dagger}_{-\sigma}a_{n -\sigma}\rangle\bigl
(1+\Sigma_{0}G_{dd\sigma}\bigr ), \label{Gamma2}
\end{eqnarray}
in which \begin{eqnarray} \langle X_{n}\rangle =\sum_m J_{dm}^\ast
\langle a^{\dagger}_{m -\sigma}a_{n-\sigma}\rangle. \label{X}
\end{eqnarray}
Although the first two terms on the RHS of Eq. (\ref{Gamma2}) are
similar to those in Eq. (\ref{G1}), we now need to calculate the
other two terms. Since these terms require thermal averages, which
we express using the fluctuation-dissipation Eq. (\ref{FD}), they
end up with integrals which involve $G_{dd-\sigma}$, leading
finally to our integral equation for the dot GF. The details of
these calculations are presented in Appendix \ref{App2}.

Inserting the results for the necessary thermal averages into the
RHS of Eq. (\ref{Gamma2}), the results are used in Eq. (\ref{XXX})
for the combination $-\sum_n J_{dn} \Gamma_{n\sigma}^{(2)}$ which
is needed in the RHS of Eq. (\ref{eomgamma}). Adding also the
corresponding combination
 for $\Gamma^{(1)}$ [Eq.
(\ref{g1sof})],  yields our final result for $\Gamma_\sigma$. It
remains to insert it back into the EOM (\ref{eomG0}) for the dot
Green function $G_{dd\sigma}$. This yields an integral equation
for this function,
\end{multicols}
\begin{eqnarray}
(\omega -\epsilon_{d}-\Sigma_0)G_{dd\sigma}&=&1-\langle
n_{d-\sigma}\rangle +G_{dd\sigma}\int\Bigl (-\frac{d\omega '}{2\pi
i}\Bigr )\frac{f(\omega ')}{\omega -\omega '+i\eta}\Bigl
[(1+\Sigma '_0 G'_{dd-\sigma})(\Sigma '_0-\Sigma _0)-(\omega
'\rightarrow\omega '^{\ast})\Bigr
]\nonumber\\
&+&(1+\Sigma_0 G_{dd\sigma})\int\Bigl (-\frac{d\omega '}{2\pi
i}\Bigr )\frac{f(\omega ')}{\omega -\omega '+i\eta}\Bigl
[G'_{dd-\sigma}(\Sigma_0-\Sigma '_0)-(\omega '\rightarrow\omega
'^{\ast})\Bigr ],\ \ \ \eta\rightarrow 0^{+}.\label{Gsof1}
\end{eqnarray}
Here and below, $G_{dd\sigma}$ and $\Sigma_0$ are understood to be
functions of $\omega \rightarrow \omega+i\eta$, and the primes
denote a dependence on $\omega '\rightarrow \omega'+i\eta$, with
$\omega '^{\ast}\rightarrow \omega '-i\eta$. Two comments are in
place here. First, for the simplest SIAM, when $\Sigma_0(\omega)
\approx -i\Delta_b$  is independent of $\omega$, we have
$\Sigma_0(\omega^\ast) \approx i\Delta_b$, so that only the parts
with $\omega'^\ast$ survive. Furthermore, in this case one can
factorize $\Delta_b$ out of the integrals. This reproduces the
integral equation of Lacroix, \cite{lacroix}
\begin{eqnarray}
(\omega-\epsilon_d+i\Delta_b)G_{dd\sigma}&=&1-\langle
n_{d-\sigma}\rangle -2i\Delta_b\Bigl [ G_{dd\sigma}\int\Bigl
(-\frac{d\omega '}{2\pi i}\Bigr )\frac{f(\omega ')}{\omega -\omega
'+i\eta}(1+i\Delta_b [G'_{dd-\sigma}]^\ast)\nonumber\\
 &-&(1-i\Delta_b
G_{dd\sigma})\int\Bigl (-\frac{d\omega '}{2\pi i}\Bigr
)\frac{f(\omega ')}{\omega -\omega
'+i\eta}[G'_{dd-\sigma}]^\ast\Bigr ].
\end{eqnarray}
\begin{multicols}{2}
\noindent However, even for this simple case there has not been an
analytic solution that covers the whole parameter range.
Second, discarding \cite{meir} correlations between the dot and
other sites on the net, e.g. $\langle d^\dagger_\sigma a_{n\sigma}
\rangle$,  amounts to neglecting the terms containing $G'_{dd}$ in
Eq. (\ref{Gsof1}). This ends up with a breakdown of the Fermi
liquid conditions at $T=0$, and with a bad approximation for
$T<T_K$.

\section{Approximate solution of the integral equation}\label{III}

We now restrict the discussion to ``non-magnetic" states, and
replace $G_{dd\uparrow}=G_{dd\downarrow}\equiv G_{dd}$.  Equation
(\ref{Gsof1}) can then be written as
\begin{eqnarray}
(\omega -\epsilon_{d}-\Sigma_{0})G_{dd} =\delta n &+&{\cal
A}G_{dd}-{\cal B}(1-i\Delta G_{dd}),\label{Gsof}
\end{eqnarray}
where $\delta n=1-\langle n_{d\uparrow}\rangle=1-\langle
n_{d\downarrow}\rangle\equiv 1- \langle n_{d}\rangle /2$ and
${\cal A}$ and ${\cal B}$ are functions of $\omega$, given by
\begin{eqnarray}
&&{\cal A}(\omega )=\int\Bigl (-\frac{d\omega '}{2\pi i}\Bigr
)\frac{f(\omega ')}{\omega -\omega '+i\eta}\nonumber\\
&&\times\Bigl [\Sigma_{0}'\Bigl
(1+G'_{dd}(\Sigma_{0}'-\Sigma_{0})\Bigr )
-(\omega '+i\eta\rightarrow\omega'-i\eta)\Bigr ],\nonumber\\
&&{\cal B}(\omega )=\int\Bigl (-\frac{d\omega '}{2\pi i}\Bigr
)\frac{f(\omega ')}{\omega -\omega '+i\eta}\nonumber\\
&&\times\Bigl [G'_{dd}(\Sigma_{0}'-\Sigma_{0})-(\omega '+i\eta
\rightarrow\omega'-i\eta)\Bigr ].\label{ab}
\end{eqnarray}

For reasons that will be explained later, it is now convenient to
replace $f(\omega')$ inside the integrals by
$[f(\omega')-1/2]+1/2$. The part related to the constant 1/2 can
then be calculated using the Kramers-Kronig relations,
\begin{eqnarray}
\int\Bigl (-\frac{d\omega '}{2\pi i}\Bigr )\frac{1}{\omega -\omega
'+i\eta}\bigl (F(\omega')-F(\omega')^\ast\bigr ) \equiv F(\omega),
\end{eqnarray}
and thus we find
\begin{eqnarray}
&&{\cal A}(\omega )=\frac{1}{2}\Sigma_0(\omega)+\int\Bigl
(-\frac{d\omega '}{2\pi i}\Bigr
)\frac{f(\omega ')-1/2}{\omega -\omega '+i\eta}\nonumber\\
&&\times\Bigl [\Sigma_{0}'\Bigl
(1+G'_{dd}(\Sigma_{0}'-\Sigma_{0})\Bigr )
-(\omega '+i\eta\rightarrow\omega'-i\eta)\Bigr ],\nonumber\\
&&{\cal B}(\omega )=\int\Bigl (-\frac{d\omega '}{2\pi i}\Bigr
)\frac{f(\omega ')-1/2}{\omega -\omega '+i\eta}\nonumber\\
&&\times\Bigl [G'_{dd}(\Sigma_{0}'-\Sigma_{0})-(\omega '+i\eta
\rightarrow\omega'-i\eta)\Bigr ].\label{abnew}
\end{eqnarray}
We next follow Lacroix, \cite{lacroix} and assume
 that the
integrals are dominated by the region $\omega '\simeq \omega$,
namely that
\begin{eqnarray}
\int d\omega ' F(\omega ,\omega ')\frac{f(\omega ')-1/2}{\omega
-\omega '+i\eta }\simeq F(\omega ,\omega ){\rm X}(\omega
),\label{lacapp}
\end{eqnarray}
and thus
 \begin{eqnarray} {\cal B}(\omega ) \approx \Delta(\omega)
G^{\ast}_{dd}(\omega ){\rm X}(\omega ),
\label{Bapprox}\end{eqnarray}
  where $\Delta(\omega)=-\Im \Sigma_0(\omega)$ and
\cite{bloomfield}
\begin{eqnarray}
{\rm X}(\omega )&=&\int_{-D}^{D}\frac{d\omega
'}{\pi}\frac{f(\omega ')-1/2}{\omega -\omega '+i\eta}\nonumber\\
&=& \frac{1}{\pi}\Bigl
[\frac{1}{2}\ln\frac{\beta^{2}(D^{2}-\omega^{2})}{(2\pi )^{2}
}-\Psi \bigl (\frac{1}{2}+\frac{\beta\omega}{2\pi i}\bigr )\Bigr
],
\end{eqnarray}
($\Psi $ is the Digamma function). A similar procedure yields
\begin{eqnarray}
{\cal A}(\omega ) \approx \Delta(\omega) \bigl
[-\frac{i}{2}+X(\omega )(1+i\Delta(\omega) G^{\ast}(\omega ))\bigr
]. \label{Aapprox}\end{eqnarray}

The reason for using the above transformation is that when one
analyzes the equivalent equations for finite $U$, then one
requires particle hole symmetry, namely the relation
\begin{eqnarray}
\widetilde{G}_{dd}(\omega )=-G_{dd}^{\ast}(-\omega ),
\end{eqnarray}
where the `tilde' denotes particle-hole transformed quantities. In
that case, this relation is equivalent to the relations
\begin{eqnarray}
&&\widetilde{{\cal B}}(\omega )=-{\cal B}^{\ast}(-\omega
),\nonumber\\
&&\widetilde{{\cal A}}(\omega )={\cal A}^{\ast}(-\omega
)-\Sigma_{0}^{\ast}(-\omega ). \end{eqnarray} These relations
should hold also for any approximate solution. \cite{dworin} It is
easy to check that these relations are obeyed by our approximate
expressions (\ref{Bapprox}) and (\ref{Aapprox}), but would not
hold if we made the approximation (\ref{lacapp}) on the original
equations, before shifting $f(\omega')$ by 1/2.
\cite{dworin,lacroix}

Taking $\delta n$ as a parameter, and defining the scaled variable
\begin{eqnarray}
z(\omega) \equiv
\frac{\omega-\epsilon_d-\delta\epsilon_d(\omega)}{2\Delta(\omega)},
\label{zzz}
\end{eqnarray} the solution of Eq. (\ref{Gsof}) can be written as
\begin{eqnarray}
G_{dd}(\omega )=g(\omega )\Bigl (\delta n+i{\rm Q}(\omega )/{\rm
X}^{\ast}(\omega )\Bigr ), \label{APPRG}
\end{eqnarray}
with \begin{eqnarray} g&=&\frac{1}{\omega
-\epsilon_{d}-\delta\epsilon_{d}+
i3\Delta /2}\equiv\frac{1}{2\Delta(z+3i/4)},\label{gg}\\
{\rm Q}&=&{\rm S}-[{\rm S}^{2}+|{\rm X}|^{2}(\frac{3}{2}\delta
n-\delta n^{2})]^{1/2},\label{QQ}\\  {\rm
S}&=&z^2+\frac{9}{16}-z\Re X+(\delta n-\frac{3}{4})\Im X,
\label{SS}
\end{eqnarray}
and we have omitted the explicit dependence on $\omega$ for
brevity. For some purposes it is more convenient to write Eq.
(\ref{APPRG}) as
\begin{eqnarray}
G_{dd}(\omega )=\frac{\frac{3}{2}}{1-{\cal C}(\omega
)}\frac{1}{\omega -\bigl (\epsilon_{d}+\delta E_{d}(\omega )\bigr
)+i\frac{3}{2}\Delta(\omega) },\nonumber\\ \label{GGG}
\end{eqnarray}
where the real functions ${\cal C}(\omega )$ and $\delta
E_{d}(\omega )$, which result from the strong interactions on the
dot, depend on $T$ and on $\Sigma_{0}(\omega)$ and are given by
\begin{eqnarray}
&&\delta E_{d}(\omega )=\delta\epsilon_{d}-2\Delta (\Re {\rm
X}){\cal C}/(1-{\cal
C}),\nonumber\\
&&{\cal C}(\omega )=\frac{1}{2\Delta^{2}Q\big |g(i+\delta n{\rm
X}^{\ast}/Q)\big |^{2}}\equiv\frac{2(z^2+9/16)}{Q|i+\delta n{\rm
X}^{\ast}/Q|^2}.\label{mm}
\end{eqnarray}
Equation (\ref{GGG}) forms our general result. We emphasize again
that all we need to know is the non-interacting self-energy,
$\Sigma_0(\omega)$.

As seen explicitly from the above equations, our analytic
expression for $G_{dd}$ depends on the parameter $\delta
n=1-\langle n_d \rangle/2$, which needs to be determined
self-consistently, via Eq. (\ref{selfcon}). We discuss this
condition in a specific example below.

\section{The Fermi liquid relations and the Kondo temperature}
\label{fermi}

At $T=0$, one easily sees that \begin{eqnarray} \Re{\rm X}(\omega)
\approx \ln |D/\omega|/\pi, \end{eqnarray}
 diverging at the Fermi
energy $\omega \rightarrow 0$. At $T=\omega=0$ we thus have ${\cal
C}\rightarrow 0$ and ${\cal C}\Re X \rightarrow u(z)$, with
\begin{eqnarray}
u(z)=\frac{2(z^2+9/16)[z+\sqrt{z^2+\delta n(1.5-\delta
n)}]}{[z+\sqrt{z^2+\delta n(1.5-\delta n)}]^2+\delta n^2},
\end{eqnarray}
and
$z=z(\omega=0)=-[\epsilon_s+\delta\epsilon_d(0)]/[2\Delta(0)]$.
Therefore,
\begin{eqnarray} G_{dd}(0)=\frac{1}{\Delta(0) \bigl
((4/3)[z-u(z)]+i\bigr )}.\label{GT0}
\end{eqnarray}
For a very deep level on the dot, $z \gg 1$, $u(z)$ approaches
$z$, and $G_{dd}$ approaches $-i/\Delta(0)$, ending up with  high
plateaus in the LDOS and in
the phase of the complex $G_{dd}(\omega)$ sticking to $\pi/2$. All
of these characteristics, which are hallmarks of the Kondo
behavior, \cite{oreg} are observed in the examples below. As
$\epsilon_d$ moves up from large negative values, Eq. (\ref{GT0})
shows a gradual crossover away from this ``unitary" limit, and
$G_{dd}$ becomes mainly real and small as $z \rightarrow -\infty$.

Eq. (\ref{GT0}) also ensures that the GF obeys the Fermi-liquid
relations \cite{bickers1} at $T=0$:  (i) The imaginary part of the
SE on the Fermi level is the same as in the absence of
interactions, i.e. $-\Delta(0)$, confirming the unitarity limit;
\cite{langreth,ng} (ii) For $\epsilon_d \ll -\Delta$, the LDOS
there [i.e., $\rho_{d}(\omega=0)\equiv \rho_{d\sigma}(0)$]
approaches the finite value $1/[\pi\Delta(0)]$. We emphasize that
these results apply for {\it any} complex network; the only input
is $\Sigma_0(\omega)$.

The Kondo effect concerns the behavior of the density of states
near the Fermi energy at low temperature.   In that regime,
$G_{dd}$ is dominated by its imaginary part, and thus
\begin{eqnarray} \rho_{d} \approx 1/[\pi \Delta(\omega)(1-{\cal
C})].
\end{eqnarray}
For $\Re X(\omega)
> z(\omega) \gg 1$, rne has $S\approx z(z-\Re X)<0$, $Q\approx 2S$ and the
leading $\omega$-dependence of ${\cal C}$ is
\begin{eqnarray}
&&{\cal C} \sim \frac{z}{z-\Re X}.\label{asym}
\end{eqnarray}
We thus end up with a logarithmic cusp in $\rho_{d}$ at
$\omega=0$. The related narrow peak reaches one half of its peak
value (equal to $1/[\pi\Delta(0)]$) when ${\cal C}\approx -1$. One
definition of the Kondo energy $T_K$ is to identify $T_K$ with the
half-width of this peak. Solving ${\cal C}(\omega=\pm T_K)=-1$ and
using Eq. (\ref{asym}) thus yields $\Re X=2z$, i.e.
\begin{eqnarray}
T_K(\epsilon_d)=D \exp[-a \pi
|\epsilon_d+\delta\epsilon_d(0)|/\Delta(0)], \label{Tkondo}
\end{eqnarray}
with $a=1$. This result, which agrees with that of Lacroix,
\cite{lacroix} is qualitatively similar but quantitatively
different from the presumably exact $T_K$ as given by Haldane,
\cite{haldane} which has $a=1/2$. However, $T_K$ only represents
some crossover energy scale, and we expect the above solution to
follow the qualitative variations of $T_K$ with the system
parameters, which are contained in $\delta\epsilon_d$ and in
$\Delta$. It should be emphasized that $T_K$ depends on {\it both}
$\Delta$ {\it and} $\delta\epsilon_d$ (some authors ignore the
part coming from the real part of the SE, $\delta\epsilon_d$). For
example, for the Aharonov-Bohm interferometer $T_K$ oscillates
with the flux. \cite{more}

At the Fermi energy, $\omega=0$, and for fixed $\epsilon_d<0$, one
expects $\rho_{d}(0)$ to decrease with increasing $T$. In this
limit, $X(\omega=0)= \ln|AD/T|/\pi$, with $A=e^{-\Psi(1/2)}/(2\pi)
\approx 1.13387$. For $|\epsilon_d| \gg \Delta(0)$ we then find
the asymptotic relation
\begin{eqnarray}
&&\rho_{d}(0) \approx \frac{1}{\pi \Delta(0)(1-{\cal
C})}\sim\frac{\Re X-z}{\pi \Delta(0)\Re X}\nonumber\\
 &&=
\frac{1}{\pi\Delta(0)}\Bigl
(1+\frac{\pi(\epsilon_d+\delta\epsilon_d)}{2\Delta(0)\ln(AD/T)}\Bigr
).\label{rhoT}
\end{eqnarray}
This density of states decreases to one half of its maximum $T=0$
value at a temperature $A T_K$, with $T_K$ given in Eq.
(\ref{Tkondo}). Although the result (\ref{rhoT}) is qualitatively
correct, and is also consistent with our $T_K$, its explicit
low-temperature dependence disagrees with exact expectations: we
obtain a logarithmic behavior, whereas Fermi liquid theory
predicts a $T^2$ dependence at low $T$. \cite{nozieres,costi} This
is an artifact of our approximation. However, at high $T$ we do
recover the usual logarithmic variation.

In the opposite limit, of $|z|\gg \Re X \sim 1$, one finds that
$|Q/X| \ll 1$, and  Eq. (\ref{APPRG}) implies that $G_{dd}(\omega
)$ approaches $\delta n~g(\omega )$. The appearance of $\delta n$
in the numerator, in place of 1, results from the fact that some
weight of the spectral function is pushed to infinite $U$.
\cite{meir} Another interesting point concerns the factor $3/2$ in
Eqs. (\ref{gg}) and (\ref{GGG}). This factor may be explained
heuristically \cite{ralph} by the fact that while both spin
directions are accessible for tunnelling into the dot, only a
single one can tunnel out of it.

\section{Example 1: the simple SIAM}\label{IV}

In the following two examples, we assume simple semi-infinite
one-dimensional leads, with identical nearest-neighbor hopping
matrix elements, $-J$, and with a lattice constant $a$. The
eigenenergies of each ${\cal H}_L$ are therefore $\epsilon_k=-2J
\cos ka$, with eigenfunctions $\langle n|k\rangle=\sin
nka\sqrt{2/\Omega}$ ($\Omega \rightarrow \infty$ is the length of
the lead), and $D=2J$.  We also assume that the leads $L$ and $R$
are attached to the mesoscopic network only at one site, with
hopping matrix elements $J_\ell$ and $J_r$. For the simple SIAM,
when the two leads are directly connected to the dot, we show in
Appendix \ref{A} that the leads generate a self-energy
\begin{eqnarray}
\Sigma_0(\omega)=-e^{i|q|a}\bigl (|J_\ell|^2+|J_r|^2\bigr )/J
\equiv-e^{i|q|a}\Delta_b
\end{eqnarray}
 on the dot, with $q$ determined by $\omega=-2J \cos qa$ and with
\begin{eqnarray}
 \Delta_b \equiv \Gamma_\ell+\Gamma_r.\label{delta}
\end{eqnarray}
Here we have used the notation
\begin{eqnarray}
\Gamma_{\ell,r} \equiv |J_{\ell,r}|^2/J =\pi {\cal
N}(0)|J_{\ell,r}|^2,
\end{eqnarray}
with ${\cal N}(\omega)$ being the density of the band states. The
width of the non-interacting resonance at $\epsilon_d$ is thus
equal to $\Delta_b \sin qa$. In the figures presented below we use
the symmetric case, $\Gamma_\ell=\Gamma_r=\Delta_b/2$.
Substituting
$\Sigma_0(\omega)=\delta\epsilon_d(\omega)-i\Delta(\omega)$ into
Eqs. (\ref{APPRG})-(\ref{SS}) then yields $G_{dd}$.

\begin{figure}[htb]
\leavevmode \epsfclipon \epsfxsize=6.truecm
\centerline{\vbox{\epsfbox{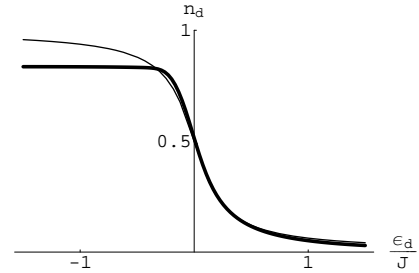}}} \vspace{0.1cm} \caption{ The
self-consistent average occupation on the dot, $\langle n_d
\rangle$ (thick line), and the Friedel added occupation $\tilde n$
at $T=0$, as function of $\epsilon_{d}$, for  the simplest SIAM
with $\Delta_b/J =0.1$. Results remain almost unchanged up to $T
\sim \Delta_b$.} \label{nd}
\end{figure}

 We start by discussing the
self-consistency condition for $\langle n_d \rangle$. Figure
\ref{nd} shows an example of our self-consistent solutions for
$\langle n_d \rangle$, at $T=0$: we start from an initial guess
for $\delta n$, then calculate $\langle n_d \rangle$ from Eq.
(\ref{selfcon}), and iterate; the procedure converges after a few
iterations.  Also shown is $\tilde n$, which represents the total
change in the electron occupation in the system due to the dot, as
determined from the Friedel phase, defined via
\cite{ng,langreth,bickers1}
\begin{eqnarray}
\tan[\frac{\pi}{2}(1-{\tilde n})] \equiv
\Re[G_{dd}(0)]/\Im[G_{dd}(0)]. \label{friedel}
\end{eqnarray}
For $\epsilon_d \ll -\Delta(0)=-\Delta_b$, $\tilde n$ approaches
1, while $\langle n_d \rangle$ remains slightly smaller. This
small difference could reflect an additional occupation of other
sites in the leads. It could also result from the inaccuracy of
our approximation for $G_{dd}(\omega )$, which becomes worse as
$\omega$ moves away from the Fermi energy; the integral in Eq.
(\ref{selfcon}) contains contributions from all $\omega$. However,
this small difference has only a small effect on the other
calculations presented below.

Although it is easy to solve for $\langle n_d \rangle$ for each
set of parameters and use the resulting $\delta n$ for other
calculations, the qualitative results are only weakly affected if
one uses an arbitrary smooth variation of $\langle n_d \rangle$
from 0 to 1 as $\epsilon_d$ varies from $+\infty$ to $-\infty$. In
the calculations below, we used such a smooth variation for
$\delta n$, with a width of order $\Delta_b$.

An alternative approximate expression for $\delta n$ follows from
Eq. (\ref{APPRG}). It turns out that for most of the integration
range in Eq. (\ref{selfcon}), $G_{dd}(\omega)$ is dominated by the
first term there. Neglecting the second term, one has $1-\delta
n=\langle n_{d\sigma}\rangle\approx\delta n~ n_0$, with $n_0=-\int
d\omega \Im [g(\omega)]/\pi$ depending  only on the
non-interacting parameters. Thus, $\delta n \approx 1/(1+n_0)$.
For $\Delta_b$ between 0.1 and 0.01 and for $\epsilon_d<\Delta_b$,
we find this estimate to be within $\sim 10\%$ from the full
self-consistent value.

\begin{figure}[htb]
\leavevmode \epsfclipon \epsfxsize=6.truecm
\centerline{\vbox{\epsfbox{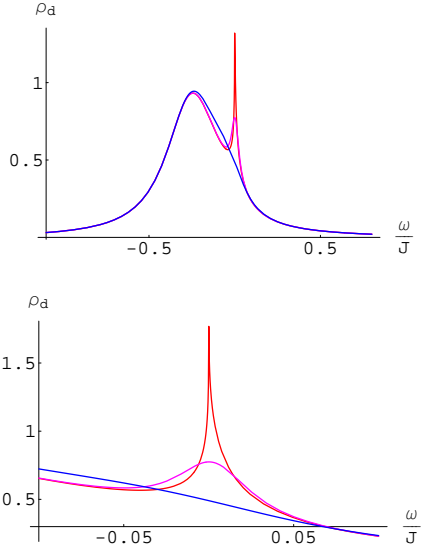}}} \caption{ The LDOS on the
dot, for the simple SIAM, with $\epsilon_d=-.3,~\Delta_b=.1$ and
$T=0,~.01,~.05$ (all energies are in units of $J$). The Kondo peak
values near $\omega=0$, shown with more detail in the lower panel,
decrease with increasing $T$. } \label{DOST}
\end{figure}

\begin{figure}[htb]
\leavevmode \epsfclipon \epsfxsize=6.truecm
\centerline{\vbox{\epsfbox{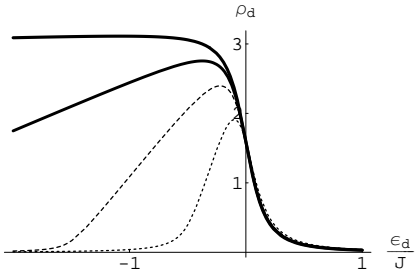}}} \vspace{0.1cm} \caption{
The LDOS at the Fermi energy, $\rho_d(0)$,  versus $\epsilon_{d}$,
for $T=0,~10^{-30},~10^{-10},~10^{-3}$ and $\Delta_b=.1$;
$\rho_{d}(0)$ increases with decreasing $T$.} \label{DOSED}
\end{figure}

Figure \ref{DOST} presents the LDOS [Eq. (\ref{rho})] versus
$\omega$, for the simple SIAM, with the parameters as indicated.
Plots at lower $T$ are indistinguishable from the one shown at
$T=.01J$. Our approximation reproduces the Kondo peak at low $T$,
in addition to the much broader peak at $\epsilon_d$. Figure
\ref{DOSED} shows $\rho_d$  at the Fermi energy ($\omega=0$)
versus $\epsilon_d$. Note the very slow (logarithmic) increase of
$\rho_{d\sigma}$ with decreasing $T$, in agreement with Eq.
(\ref{rhoT}).

We now turn to the conductance, ${\cal G}$. For the simple SIAM,
one has \cite{ng,meir} \begin{eqnarray} {\tilde{\cal G}}(\omega)=
-\frac{4\Gamma_\ell\Gamma_r}{\Gamma_\ell+\Gamma_r} \Im
G_{dd}(\omega).\label{con1}
 \end{eqnarray} Equation (\ref{GGG})
implies that
\begin{eqnarray}
\Im G_{dd}=-(1-{\cal C})\Delta_b|G_{dd}|^2.
\end{eqnarray}
 For $T=0$, when ${\cal C}\rightarrow 0$, we also have ${\tilde{\cal
G}}(\omega)=4\Gamma_\ell\Gamma_r |G_{dd}|^2$. At finite
temperatures, these two expressions for the conductance exhibit
the same qualitative behavior as $\rho_d$ (Fig. \ref{DOSED}). The
quantitative difference between the two expressions, which
increases with $T$ and with large negative $\epsilon_d$,
represents a breakdown of unitarity which may be an artifact of
our approximation.

At low $T$, $(-\partial f/\partial \omega)$ is practically a delta
function, and Eq. (\ref{cond}) yields ${\cal G}\approx
(2e^2/h){\tilde{\cal G}}(0)$, exhibiting the qualitative behavior
shown in Fig. \ref{DOSED}. At higher $T$, the peak in ${\cal G}$
slightly below $\epsilon_d=0$ becomes lower and broader than that
of $\rho_{d}(\omega=0)$. At fixed $\epsilon_d <0$, increasing $T$
results with a decreasing ${\cal G}$, but with an interesting
superimposed peak at $T={\cal O}(|\epsilon_d|$) (when the peak at
$\epsilon_d$ starts to contribute).

\begin{figure}[htb]
\leavevmode \epsfclipon \epsfxsize=6.truecm
\centerline{\vbox{\epsfbox{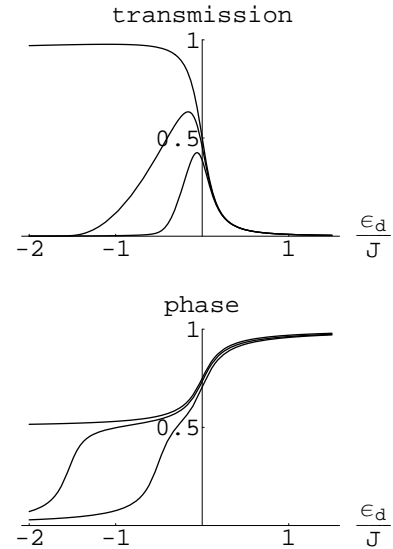}}} \caption{"Transmission" and
``transmission phase" (in units of $\pi$) for the simple SIAM, for
$\Delta_b=0.1$ and for $T=0,~10^{-10},~10^{-3}$.} \label{phase}
\end{figure}

Some experiments \cite{ji} place the QD on one branch of an open
Aharonov-Bohm interferometer, and attempt to extract the {\it
transmission phase} for scattering through the dot.  The
transmission phase is usually related to the phase of
$G_{dd}$.\cite{oreg} Without getting into the question of what is
really measured in the interferometer, \cite{bih} it is still of
interest to study the latter phase. Figure \ref{phase} thus shows
the ``transmission", represented by $\Delta_b^2|G_{dd}|^2$, and
the ``transmission phase", represented by [see Eq.
(\ref{friedel})] $\alpha=\pi(1-\tilde n/2)$ as function of
$\epsilon_d$ for several temperatures. Interestingly, at high $T$
this phase simply grows smoothly from zero to $\pi$ through the
resonance, similarly to the non-interacting case or to the Coulomb
blockade case. However, as $T$ decreases, the peak in the
``transmission" broadens towards negative $\epsilon_d$, eventually
reaching a plateau for $T=0$ (see also Fig. \ref{DOSED}). At the
same time, the phase develops an intermediate plateau at $\pi/2$.
\cite{oreg} This plateau begins at an energy
 $\epsilon_d$ which is roughly given by  $T \sim T_K(\epsilon_d)$,
 as defined in Eq. (\ref{Tkondo}). Studying the energy
where this phase grows from zero to $\pi/2$ thus suggests another
way to define the crossover temperature $T_K$.

\section{Example 2: the QD on a ``T" network}\label{T}

Our equations also become very simple for the ``T" network, when
the dot sits on a side branch. Such a network has recently
attracted both theoretical \cite{Ttheor,jps} and experimental
\cite{Texp} interest, as the simplest realization of the
Fano-Kondo effect. The mesoscopic network now consists of a single
site ``0". This site has one bond connected to the dot, with
hopping $J_x$, and two bonds connected to two leads, with hopping
$J_\ell$ and $J_r$. In this case, the matrix $F_{11}$ in Appendix
\ref{A} is of order $1 \times 1$, and Eqs. (\ref{DD})
 and (\ref{AAA}) yield  ${\cal
F}_{00}=1/[\omega-\epsilon_0+e^{i|q|a}(|J_\ell|^2+|J_r|^2)/J]$.
Using $\Delta_b$ from Eq. (\ref{delta}), we have
\begin{eqnarray}
\Sigma_0(\omega)=
\frac{|J_x|^2}{\omega-\epsilon_0+e^{i|q|a}\Delta_b}\equiv
\delta\epsilon_d(\omega)-i\Delta(\omega).\label{sig0T}
\end{eqnarray}
In this case, ${\tilde {\cal G}}$ of Eq. (\ref{cond}) is given by
\cite{jps}
\begin{eqnarray} {\tilde{\cal
G}}(\omega)= -\frac{4\Gamma_\ell\Gamma_r}{\Gamma_\ell+\Gamma_r}
\Im G_{00}(\omega).\label{conT}
\end{eqnarray}
The calculation of $G_{00}$ follows directly from Eqs. (\ref{Gdm})
and (\ref{Glm}),
\begin{eqnarray}
G_{00}=\frac{1+\Sigma_0 G_{dd}}{\omega-\epsilon_0+i\Delta_b}.
\end{eqnarray}
Therefore, $\Im G_{00}$ is a linear combination of $\Re G_{dd}$
and of $\Im G_{dd}$, with coefficients which depend only on the
non-interacting parameters.

These equations reproduce those found e.g. in Refs.
\onlinecite{Ttheor} and \onlinecite{jps}. However, at this point
those authors use complicated numerical schemes to obtain
$G_{dd}$, which do not allow for systematic studies of the
dependence on the various parameters. In contrast, we can easily
obtain the approximate $G_{dd}$ analytically, with the same ease
as for the previous example. All we need to do is to substitute
Eq. (\ref{sig0T}) into Eqs. (\ref{APPRG})-(\ref{SS}).

\begin{figure}[htb]
\leavevmode \epsfclipon \epsfxsize=6truecm
\centerline{\vbox{\epsfbox{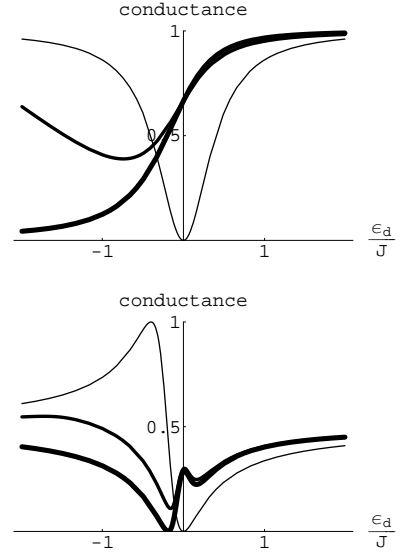}}}
\vspace{0.3cm}
 \caption{Conductance of the ``T" network, with
 $J_x=.2,~\Delta_b=.1$ (all in units of $J$).
 Thin lines: without interactions. Thick lines: with
 infinite interactions at $T=0$ (intermediate thickness
  lines: $T=10^{-5}$).
 Upper panel: $\epsilon_0=0$. Lower panel: $\epsilon_0=-.1$.}
 \label{Tcond}
 \end{figure}

For the non-interacting case, $G_{dd}$ is simply given by Eq.
(\ref{G0}). Figure \ref{Tcond} then compares the results for the
conductance (in units of $2e^2/h$), with and without interactions.
Without interactions (thin lines), for $\epsilon_0=0$ the upper
panel in Fig. \ref{Tcond} shows a symmetric Fano vanishing of the
conductance at $\epsilon_d=0$. As $\epsilon_0$ becomes more
negative (lower panel there), the conductance assumes a typical
non-symmetric Fano shape, with smaller conductances at large
$|\epsilon_d|$. The interactions (thick lines) have a negligible
effect at $\epsilon_d\gg 0$, when the dot is not occupied.
However, at $\epsilon_d<0$ and $T=0$ (thickest lines) the
interactions cause strong changes. For $\epsilon_0=0$ and large
negative $\epsilon_d$, the occupation on the dot causes the
so-called anti-Kondo effect, \cite{Ttheor} where the formation of
the Kondo singlet causes destructive interference which yields
zero conductance. For $\epsilon_0<0$, the Fano zero is shifted to
lower $\epsilon_d$, the peak disappears and the asymptotic
conductance increases. As $T$ increases (medium thick lines), the
behavior at negative $\epsilon_d$ gradually returns to that of the
non-interacting case, and the Fano zero disappears: the
conductance is always non-zero. Again, all of these phenomena are
qualitatively similar to those found in Refs. \onlinecite{Texp},
\onlinecite{Ttheor} and \onlinecite{jps}.

A slightly more complex T-network is obtained by placing another
non-interacting site, ``1", between the site 0 and the dot, with
energy $\epsilon_1$ and with hopping elements $J_x$ to both site 0
and the dot. Our mesoscopic system now contains two sites, 0 and
1, and the methods of Appendix \ref{A} yield
\begin{eqnarray}
\Sigma_0=\frac{J_x^2(\omega-\epsilon_0+e^{i|q|a}\Delta_b)}{D},
\end{eqnarray}
\begin{eqnarray}
G_{00}=\frac{\omega-\epsilon_1}{D}+\frac{J_x^4}{D^2}G_{dd},
\end{eqnarray}
with
\begin{eqnarray}
D=(\omega-\epsilon_1)(\omega-\epsilon_0+e^{i|q|a}\Delta_b)-J_x^2.
\end{eqnarray}

\begin{figure}[htb]
\leavevmode \epsfclipon \epsfxsize=6.truecm
\centerline{\vbox{\epsfbox{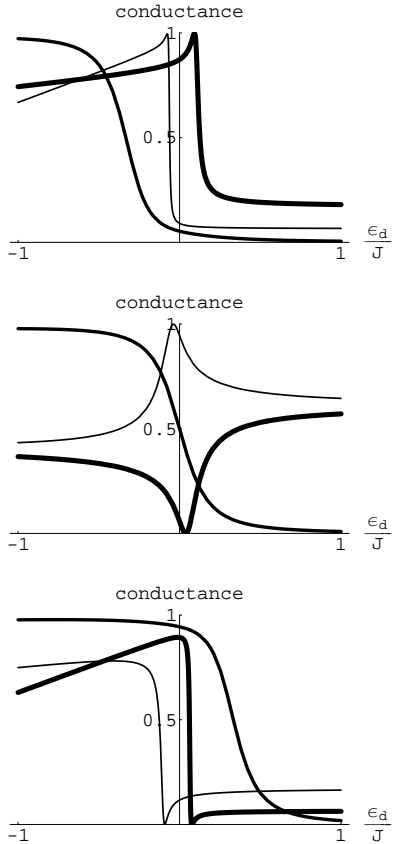}}}
 \caption{Conductance of the ``T" network, with an extra site ``1"
 between the dot and the ``intersection" site ``0".
 Plots are for
 $U=\infty,~T=0,~J_x=.2,~\Delta_b=.1$ (all in units of $J$).
 From top to bottom: $\epsilon_0=0.3,~0,~-0.3$. For each value of
 $\epsilon_0$, the figure contains graphs for $\epsilon_1=-0.5,~0,~0.5$,
 with increasing thickness.}
 \label{T11}
 \end{figure}

 Figure \ref{T11}
shows a few examples on how the interplay between $\epsilon_0$ and
$\epsilon_1$ can change the dependence of the conductance on
$\epsilon_d$ for the fully interacting case at $T=0$. For
$\epsilon_1=\epsilon_0=0$, the graph looks exactly like that for
the simple SIAM, discussed in Sec. V. Clearly, the presence of the
intermediate point turned the destructive interference into a
constructive one. Changing $\epsilon_0$ to positive (negative)
values then simply shifts the whole curve  to the left (right).
Changing $\epsilon_1$ to non-zero values generates either a "Fano"
zero in the conductance (when $\epsilon_0<0$), similar to the
results in Fig. \ref{Tcond}, or a ``Fano" resonance (when
$\epsilon_0>0$). At $\epsilon_0=0$ one observes a change from a
zero to a resonance as $\epsilon_1$ changes from positive to
negative values. All of these parameters can be easily changed
using the various gate voltages, e.g in the setup of Ref.
\onlinecite{Texp}. Without going into much further discussion, it
is clear that the intermediate point on the side branch is very
effective in changing the interference pattern between Fano-Kondo
resonances and anti-resonances. Data from such experiments can
then be used to obtain information on $G_{dd}$.

\section{Discussion}\label{V}

We have derived an approximate analytic expression for a complex
mesoscopic network, which contains a quantum dot with
electron-electron interactions, and which may connect to several
leads.  Our formulae correct and generalize earlier expressions,
and give a good qualitative interpolation between the Fermi liquid
behavior at very low temperatures and the simpler high temperature
one.

Although our formulae reproduce many features required by the
Fermi liquid theory, they are still approximate, and should thus
only be used for discussing the qualitative variation of various
quantities on the parameters characterizing the network. However,
the simplicity of our expressions allows for relatively easy
comparisons with experiments and with more complicated numerical
work.

We have demonstrated the use of our formulae for the two simple
cases of a single quantum dot attached to two leads and of the
``T" network, where the dot is connected to another site which
couples to the two leads. Indeed, we have reproduced and extended
all the expected phenomena for these two cases. In an upcoming
publication \cite{more} we shall apply this scheme to the
interesting case of the Aharonov-Bohm interferometer.

\begin{center}{\bf ACKNOWLEDGEMENTS}\end{center}

We acknowledge helpful discussions with A. Schiller, and support
to YM from the Israel Science Foundation (ISF). This project was
carried out in a center of excellence supported by the ISF (OE and
AA).

\end{multicols}
\newpage
\appendix
\begin{multicols}{2}

\section{Calculating ${\cal F}={\cal M}^{-1}$}\label{A}

We need to invert the matrix ${\cal M}=\omega+i\eta-{\cal H}_{\rm
net}$. We start by dividing our Hilbert space into two parts,
called ``1" and ``2". Then, we write
\begin{eqnarray}
{\cal M}=\left
(\begin{array}{cc}M_{11}&M_{12}\\M_{21}&M_{22}\end{array}\right ),
\end{eqnarray}
and the inverse matrix ${\cal F}$ as
\begin{eqnarray}
{\cal F}=\left
(\begin{array}{cc}F_{11}&F_{12}\\F_{21}&F_{22}\end{array}\right ).
\end{eqnarray}
It is now easy to obtain the Dyson-like equation,
\begin{eqnarray}
F_{11}=\bigl (M_{11}-M_{12}[M_{22}]^{-1}M_{21}\bigr
)^{-1}.\label{DD}
\end{eqnarray}

In our calculations we only need ${\cal F}_{mn}$ for $m,~n$ within
the finite mesoscopic part of the network (possibly including the
dot).  Identifying this part with the subspace ``1" above, we thus
only need the matrix elements of the finite matrix $F_{11}$. To
obtain these via Eq. (\ref{DD}), we need the Green function for
the leads, $[M_{22}]^{-1}$.  For our Eq. (\ref{leads}), the matrix
$M_{22}$ separates for the different leads, and we end up with
\begin{eqnarray}
F_{22}=\bigl [M_{22}-\sum_L \Sigma^{(L)}\bigr ]^{-1},\label{minv1}
\end{eqnarray}
with the self-energy matrices
\begin{eqnarray}
\Sigma^{(L)}_{mn}=-\sum_{rs}J_{mr}g^{(L)}_{rs}J_{sm}.
\end{eqnarray}
Here, the matrices $M_{22}$ and $\Sigma^{(L)}$ are all of order
$N\times N$, where $N$ is the number of sites in the mesoscopic
part of the network. Also,
\begin{eqnarray}
g^{(L)}_{rs}=\sum_k\frac{\langle r|k\rangle \langle
k|s\rangle}{\omega-\epsilon_k}
\end{eqnarray}
is the Green function for the disconnected lead $L$.

For the one-dimensional leads, we use the identity
\begin{eqnarray}
\frac{2}{\Omega}\sum_k\frac{1}{\omega+i
\eta-\epsilon_k}=\frac{1}{2iJ\sin |q|a},
\end{eqnarray}
with $\omega=-2J\cos qa$, to obtain the self-energy due to the
lead $L$, which is attached at the single point $n_L$,
\begin{eqnarray}
\Sigma^{(L)}(\omega)_{mn}=-|J_\ell|^2
e^{i|q|a}\delta_{m,n_L}\delta_{m,n_L}/J.\label{AAA}
\end{eqnarray}
 The same result also applies when the mesoscopic part is empty,
 and the leads are connected directly to the dot.

\section{Details of the EOM for $\Gamma^{(2)}$}
\label{App2}

Following the same logic that led to Eqs. (\ref{sigext}) and
(\ref{g1sof}), the formal solution to Eq. (\ref{Gamma2}) gives
\begin{eqnarray}
-\sum_n J_{dn}
\Gamma_{n\sigma}^{(2)}&=&\Sigma_0\Gamma_\sigma-\sum_{mn}J_{dm}{\cal
F}_{mn}\bigr [\langle X_{n}\rangle
G_{dd\sigma}\nonumber\\
&+&\langle d^{\dagger}_{-\sigma}a_{n -\sigma}\rangle\bigl
(1+\Sigma_{0}G_{dd\sigma}\bigr )\bigr ].\label{GG2}
\end{eqnarray}
Calculating the thermal average  $\langle
d^{\dagger}_{-\sigma}a_{n -\sigma}\rangle$ via the
fluctuation-dissipation Eq. (\ref{FD}), and using Eqs. (\ref{Gmd})
and (\ref{Gdm}), we end up with an integral which requires the sum
\begin{eqnarray}
S_1=\sum_{mn\ell}J_{dm}{\cal F}(\omega)_{m\ell}{\cal
F}(\omega')_{\ell n}J_{nd}.\label{sum1}
\end{eqnarray}
For this and for similar sums, it is helpful to use the identity
\begin{eqnarray}
{\cal F}(\omega+i\eta){\cal F}(\omega '\pm i\eta')
=\frac{ {\cal F} (\omega '\pm i\eta')-{\cal F}(\omega+i\eta)
}{\omega-\omega '+i\eta},
\end{eqnarray}
which follows from Eq. (\ref{MM}), together with a careful use of
$1/(x+i\eta)={\cal P}(1/x)-i\pi\delta(x)$.   Using Eq.
(\ref{sig0}), the sum in Eq. (\ref{sum1}) then becomes
\begin{eqnarray}
S_1=\frac{1}{\omega-\omega'+i\eta}\bigl
[\Sigma_0(\omega')-\Sigma_0(\omega)\bigr ].
\end{eqnarray}

Similar manipulations allow the calculation of the thermal
averages $\langle X_{n}\rangle$. Finally, we end up with

\end{multicols}

\begin{eqnarray}
-\sum_n J_{dn} \Gamma_{n\sigma}^{(2)}&=&\Sigma_0\Gamma_\sigma
 +G_{dd\sigma}\int\Bigl (-\frac{d\omega '}{2\pi i}\Bigr
)\frac{f(\omega ')}{\omega -\omega '+i\eta}\Bigl [(1+\Sigma
'_0G'_{dd-\sigma})(\Sigma_0-\Sigma '_0)-(\omega '\rightarrow\omega
'^{\ast})\Bigr
]\nonumber\\
&-&(1+\Sigma_0G_{dd\sigma})\int\Bigl (-\frac{d\omega '}{2\pi
i}\Bigr )\frac{f(\omega ')}{\omega -\omega '+i\eta}\Bigl
[G'_{dd-\sigma}(\Sigma_0-\Sigma '_0)-(\omega '\rightarrow\omega
'^{\ast})\Bigr ].\label{XXX}
\end{eqnarray}
As explained after Eq. (\ref{Gsof1}), $G_{dd\sigma}$ and
$\Sigma_0$ are understood to be functions of $\omega \rightarrow
\omega+i\eta$, and the primes denote a dependence on $\omega
'\rightarrow \omega'+i\eta$, with $\omega '^{\ast}\rightarrow
\omega '-i\eta$.

\begin{multicols}{2}

\end{multicols}

\end{document}